\documentstyle[multicol,aps]{revtex}
\begin{document}
\title{Scale Invariance and Lack of Self-Averaging in Fragmentation}
\author{P.~L.~Krapivsky$^\ast$, I.~Grosse$^{\ast\dag}$, and
    E.~Ben-Naim$^\ddag$} 
\address{$^\ast$Center for Polymer Studies and
    Department of Physics, Boston University, Boston, MA 02215}
\address{$^\dag$Institute for Molecular Biology and Biochemistry, Free
    University Berlin, Arnimallee 22, 14195 Berlin, Germany}
\address{$^\ddag$Theoretical Division and Center for Nonlinear Studies, Los
    Alamos National Laboratory, Los Alamos, NM 87545} 
\maketitle
\begin{abstract} 
  
  We derive exact statistical properties of a class of recursive 
  fragmentation processes. We show that introducing a fragmentation 
  probability $0<p<1$ leads to a purely algebraic size distribution in 
  one dimension, $P(x)\propto x^{-2p}$.  In $d$ dimensions, the volume 
  distribution diverges algebraically in the small fragment limit, 
  $P(V)\sim V^{-\gamma}$ with $\gamma=2p^{1/d}$. Hence, the entire 
  range of exponents allowed by mass conservation is realized.  We 
  demonstrate that this fragmentation process is non-self-averaging. 
  Specifically, the moments $Y_\alpha=\sum_i x_i^{\alpha}$ exhibit 
  significant fluctuations even in the thermodynamic limit. 

\end{abstract}
\begin{multicols}{2}
  
  Numerous physical phenomena are characterized by a set of variables,
  say $\{x_j\}$, which evolve according to a random process, and are
  subject to the conservation law $\sum_j x_j={\rm const}$.  An
  important example is fragmentation, with applications ranging from
  geology\cite{tur} and fracture\cite{solid} to the breakup of liquid
  droplets\cite{liq} and atomic nuclei\cite{nuc,red}.  Other examples
  include spin glasses\cite{sg}, where $x_j$ represents the equilibrium
  probability of finding the system in the $j^{\rm th}$ valley, genetic
  populations, where $x_j$ is the frequency of the $j^{\rm th}$ 
  allele\cite{h,djm}, and random Boolean networks\cite{kauf,flyv}.
  
  In most cases, stochasticity governs both the way in which fragments
  are produced, and the number of fragmentation events they experience.
  For example, in fragmentation of solid objects due to impact with a
  hard surface fragments may bounce several times before coming to a
  rest \cite{fragm}. The typical number of fragmentation events
  may vary greatly as it depends on the initial kinetic energy. Another
  seemingly unrelated example is DNA segmentation algorithms where
  homogeneous subsequences are produced recursively from a inhomogeneous
  sequence until a predefined homogeneity level is reached
  \cite{alg}. Here, the number of segmentation events is determined
  by the degree of homogeneity of the original sequence.
  
  In this study, we examine fragmentation models with two types of
  objects -- stable towards fragmentation and unstable.  We show that
  the size distribution is algebraic, and that the entire range of
  power-laws allowed by the underlying conservation laws can be
  realized by tuning the fragmentation probability.  Additionally,
  such processes are characterized by large sample to sample
  fluctuations, as seen from analysis of the moments of the fragment
  size distribution.
  
  Specifically, we consider the following {\em recursive fragmentation process}.
  We start with the unit interval and choose a break point $l$ in
  $[0,1]$ with a uniform probability density. Then, with probability
  $p$, the interval is divided into two fragments of lengths $l$ and
  $1-l$, while with probability $q=1-p$, the interval becomes ``frozen''
  and never fragmented again. If the interval is fragmented, we
  recursively apply the above fragmentation procedure to both of the
  resulting fragments.
  
  First, let us examine the average total number of fragments, $N$.
  With probability $q$ a single fragment is produced, and with
  probability $p$ the process is repeated with two fragments.  Hence
  $N=q+2pN$, yielding
\begin{equation}
\label{number}
N=\cases{q/(1-2p),  & if $p<1/2$;\cr
         \infty,    & if $p\geq 1/2$.\cr}
\end{equation}
The average total number of fragments becomes infinite at the critical
point $p_c=1/2$, reflecting the critical nature of the underlying
branching process\cite{Harris}.

Next, we study $P(x)$, the density of fragments of length $x$.  The
recursive nature of the process can be used to obtain the fragment
length density
\begin{equation}
\label{pxeq}
P(x)=q\delta(x-1)+2p\int_x^1 {dy\over y}\,P\left({x\over y}\right).
\end{equation}
The second term indicates that a fragment can be created only from a
larger fragment, and the $y^{-1}$ kernel reflects the uniform
fragmentation density.  Eq.~(\ref{pxeq}) can be solved by introducing
the Mellin transform
\begin{equation}
\label{mellin}
M(s) = \int dx\, x^{s-1} P(x).
\end{equation}
Eqs.~(\ref{pxeq}) and (\ref{mellin}) yield \hbox{$M(s)=q+2ps^{-1}M(s)$}
and as a result
\begin{equation}
\label{ms}
M(s)=q\left[1+{2p\over s-2p}\right].
\end{equation}
The average total number $M(1) = N$ is consistent with
Eq.~(\ref{number}), and the total fragment length $M(2)=1$ is conserved
in accord with $1=\int dx\, x P(x)$.  (Here and in the following the
integration is carried over the unit interval, i.e., $0<x<1$.)  The
inverse Mellin transform of Eq.~(\ref{ms}) gives
\begin{equation}
\label{px}
P(x)=q\left[\delta(x-1)+2p\,x^{-2p}\right].
\end{equation}
Apart from the obvious $\delta$-function, the length density is a purely
algebraic function. In particular, the fragment distribution diverges
algebraically in the limit of small fragments. Given such an algebraic
divergence near the origin $P(x)\sim x^{-\gamma}$, length conservation
restricts the exponent range to $\gamma<2$. In our case $\gamma=2p$, and
since $0<p<1$, the entire range of acceptable exponents emerges by
tuning the only control parameter $p$.

Interestingly, at the critical point $p_c={1\over 2}$, the fragment
length distribution becomes independent of the initial interval length.
Starting from an interval of length $L$, Eq.~(\ref{px}) can be
generalized to yield
\begin{equation}
P(x)=q\delta(x-L)+2pq L^{1-2p}x^{-2p}.
\end{equation}
Thus, the critical point may be detected by observing the point at which
the segment distribution becomes independent of the original interval
length.
 
The recursive fragmentation process can be generalized to $d$
dimensions.  For instance, in two dimensions we start with the unit
square, choose a point $(x_1, x_2)$ with a uniform probability density,
and divide, with probability $p$, the original square into four
rectangles of sizes $x_1 \times x_2$, $x_1 \times (1-x_2)$, $(1-x_1)
\times x_2$, and $(1-x_1) \times (1-x_2)$.  With probability $q$, the
square becomes frozen and we never again attempt to fragment it. The
process is repeated recursively whenever a new fragment is produced.

Let $P({\bf x})$, ${\bf x}\equiv (x_1,\ldots,x_d)$, be the probability
density of fragments of size $x_1\times \cdots \times x_d$. This
quantity satisfies
\begin{equation}
\label{pxdeq}
P({\bf x})= q\delta({\bf x}-{\bf 1})
 + 2^dp \int {d{\bf y}\over y_1\cdots y_d}
P\left({x_1\over y_1},\ldots,{x_d\over y_d}\right)\!.\!\!
\end{equation}
with $\int d{\bf y}=\int dy_1\cdots \int dy_d$. Following the steps
leading to Eq.~(\ref{ms}), we find that the $d$-dimensional Mellin
transform, defined by \hbox{$M({\bf s})= \int d{\bf x}\,
  x_1^{s_1-1}\cdots x_d^{s_d-1} P({\bf x})$} with the shorthand notation
\hbox{${\bf s}\equiv (s_1,\ldots,s_d)$} obeys
\begin{equation}
\label{msd}
M({\bf s})=q\left[1+{\alpha^d\over s_1\cdots s_d-\alpha^d}\right], 
{\quad} {\rm with} {\quad} \alpha = 2 p^{1/d}.
\end{equation}
Eq.~(\ref{msd}) gives the total average number of fragments, $N=M({\bf
  1})=q/(1-2^dp)$ if $p<2^{-d}$ and $N=\infty$ if $p \geq 2^{-d}$.  One
can also verify that the total volume $M({\bf 2})=1$ is conserved.
Interestingly, there is an additional infinite set of conserved
quantities: all moments whose indices belong to the hyper-surface
$s_1^*\cdots s_d^*=2^d$ satisfy $M({\bf s}^*)=1$. In a continuous time
formulation of this process the same moments were found to be integrals
of motion \cite{kb,rh,btv}.  The existence of an infinite number of
conservation laws is surprising, because only the volume conservation
has a clear physical justification.

Next, we study the volume density $P(V)$, defined by
\begin{equation}
P(V)=\int d{\bf x}\, P({\bf x})\,\delta\left(V-x_1\cdots x_d\right).  
\end{equation}
The Mellin transform \hbox{$M(s) = \int dV V^{s-1} P(V)$} can be
obtained from Eq.~(\ref{msd}) by setting $s_i=s$,
\begin{equation}
M(s)=q\left[1+{\alpha^d \over s^d-\alpha^d}\right].
\end{equation}
Using the $d^{\rm th}$ root of unity, $\zeta = e^{2\pi i/d}$, and the
identity ${1\over s^d-1}={1\over d}\sum_{k=0}^{d-1} {\zeta^k \over
  s-\zeta^k}$, $M(s)$ can be expressed as a sum over simple poles at
$\alpha \zeta^k$.  Consequently, the inverse Mellin transform is given
by a linear combination of $d$ power laws
\begin{equation}
\label{pv}
P(V)=q\left[\delta(V-1)+
{\alpha \over d}\sum_{k=0}^{d-1} \zeta^kV^{-\alpha\zeta^k}\right].
\end{equation}
One can verify that this expression equals its complex conjugate and
hence, it is real.  Additionally, the one-dimensional case (\ref{px}) is
recovered by setting $d=1$.
 
The small-volume tail of the distribution can be obtained by noting that
the sum in Eq.~(\ref{pv}) is dominated by the first term in the series,
which leads to
\begin{equation}
P(V)\simeq A V^{-\gamma} \qquad {\mbox{as}} \qquad V\to 0,
\end{equation}
with $\gamma=\alpha=2p^{1/d}$ and $A = \alpha q/d$. Although the value
of the exponent changes, the possible range of exponents for this
process remains the same since $0<2p^{1/d}<2$ when $0<p<1$. In the
infinite dimension limit, $P(V)$ becomes universal: $P(V)\sim V^{-2}$.

The leading behavior of $P(V)$ in the large size limit can be derived by
using the Taylor expansion and the identity $\sum_{k=0}^{d-1}
\zeta^{kn}=\delta_{n,0}$ for $n=0,\ldots,d-1$. One finds that in higher
dimensions the volume distribution vanishes algebraically near its
maximum value,
\begin{equation}
P(V)\simeq B_d (1-V)^{d-1} \qquad {\mbox{as}} \qquad V\to 1,
\end{equation} 
with $B_d = \alpha^d/(d-1)!$.

In fact, the entire multivariate fragment length density can be obtained
explicitly. This can be achieved by expanding the geometric series
\begin{eqnarray*}
{\alpha^d\over  s_1\cdots s_d-\alpha^d} = \sum_{n\geq 0}
\prod_{i=1}^d \left({\alpha \over s_i}\right)^{n+1},
\end{eqnarray*}
and performing the inverse Mellin transform for each variable
separately.  Using the transform \hbox{$\int dx\,x^{s-1}\left[\ln
    {1\over x}\right]^{n} =n!s^{-n-1}$} gives
\begin{equation}
\label{psd} 
P({\bf x})=q\left[\delta({\bf x}-{\bf 1})+\alpha^d F_d(z)\right],
\end{equation}
with the shorthand notations
\begin{eqnarray}
\label{fd} 
F_d(z)=\sum_{n=0}^\infty \left(z^n\over n!\right)^d, 
\qquad
z = \alpha \left(\prod_{i=1}^d \ln{1\over x_i}\right)^{1/d}.
\end{eqnarray}
In two dimensions, $F_2(z)=I_0(2 z)$ where $I_0$ is the modified Bessel
function.

The small size behavior of $P({\bf x})$ can be obtained by using the
steepest decent method. The leading tail behavior, \hbox{$F_d(z)\simeq
  (2\pi z)^{1-d\over 2} e^{z d}$} for $z\gg 1$, corresponds to the case
when at least one of the lengths is small, i.~e.~$x_i\ll 1$.  Returning
to the original variables we see that the fragment distribution exhibits
an unusual ``log-stretched-exponential'' behavior
\begin{eqnarray}
P({\bf x})\sim 
\left[\prod_{i=1}^d \ln {1\over x_i}\right]^{-{d-1\over 2d}} 
\!\!\!\!\!
\exp\left[d\alpha\left(\prod_{i=1}^d \ln {1\over x_i}\right)^{1/d}\right].
\end{eqnarray}

The fragment distribution represents an average over infinitely many
realizations of the fragmentation process; hence, it does not capture
sample to sample fluctuations.  These fluctuations are important in
non-self-averaging systems, where they do not vanish in the
thermodynamic limit.  Useful quantities for characterizing such
fluctuations are the moments $Y_{\alpha}$\cite{der,df}
\begin{eqnarray}
\label{defY}
Y_\alpha=\sum_i x_i^\alpha,
\end{eqnarray}
where the sum runs over all fragments.

We are interested in the average values $\langle Y_\alpha \rangle$ and
$\langle Y_\alpha Y_\beta\rangle$.  For integer $\alpha$, $\langle
Y_\alpha \rangle$ is the probability that $\alpha$ points randomly
chosen in the unit interval belong to the same fragment.  The expected
value of $Y_\alpha$ satisfies
\begin{equation} 
\label{Yav}
\langle Y_\alpha \rangle=q+p\langle Y_\alpha \rangle 
\int dy \left[y^\alpha+(1-y)^\alpha\right].
\end{equation}
The first term corresponds to the case where the unit interval is not
fragmented, and the second term describes the situation when at least
one fragmentation event has occurred. Eq.~(\ref{Yav}) gives
\begin{equation}
\label{Yaver}
\langle Y_\alpha \rangle=q\left[1+{2p\over \alpha+1-2p}\right]
\end{equation}
if $\alpha>2p-1$, and $\langle Y_\alpha \rangle=\infty$ if $\alpha\leq
2p-1$.  As expected, Eq.~(\ref{Yaver}) agrees with the moments of $P(x)$
obtained by integrating Eq.~(\ref{px}), $\langle Y_{\alpha}\rangle=\int
dx\, x^{\alpha} P(x)$.

However, higher order averages such as $\langle Y_\alpha Y_\beta\rangle$
do not follow directly from the fragment density.  For integer $\alpha$
and $\beta$, $\langle Y_\alpha Y_\beta\rangle$ is the probability that,
if $\alpha+\beta$ points are chosen at random, the first $\alpha$ points
all lie on the same fragment, and the last $\beta$ points all lie on
another (possibly the same) fragment.  This quantity satisfies
\begin{eqnarray}
\langle Y_\alpha Y_\beta\rangle&=&q+p\langle Y_\alpha Y_\beta\rangle 
\int dy\left[y^{\alpha+\beta}+(1-y)^{\alpha+\beta}\right]\\
&+&p\langle Y_\alpha \rangle \langle Y_\beta \rangle \int dy
\left[y^\alpha(1-y)^\beta+(1-y)^\alpha y^\beta\right],\nonumber
\end{eqnarray}
yielding
\begin{eqnarray}
\label{yab}
\langle Y_\alpha Y_\beta\rangle&=&q+{2pq\over \alpha+\beta+1-2p}\\
&+&2p\,{\Gamma(\alpha+1)\Gamma(\beta+1)\over \Gamma(\alpha+\beta+1)}\,
{\langle Y_\alpha \rangle \langle Y_\beta\rangle
\over \alpha+\beta+1-2p}\nonumber
\end{eqnarray} 
when $\alpha,\beta,\alpha+\beta>2p-1$, and $\langle Y_\alpha
Y_\beta\rangle =\infty$ otherwise.

Eq.~(\ref{yab}) shows that $\langle Y_\alpha Y_\beta\rangle\neq\langle
Y_\alpha\rangle\langle Y_\beta \rangle$, and in particular, $\langle
Y^2_\alpha \rangle\neq\langle Y_\alpha\rangle^2$. Therefore,
fluctuations in $Y_\alpha$ do not vanish in the thermodynamic limit, and
the recursive fragmentation process is non-self-averaging.  While for
$p<1/2$ non-self-averaging behavior is expected because the average
number of fragments is finite, the emergence of non-self-averaging for
$p>1/2$ is surprising. Hence, statistical properties obtained by
averaging over all realizations are insufficient to probe sample to
sample fluctuations.

In principle, higher order averages such as $\langle Y_\alpha^n\rangle$
can be calculated recursively by the procedure outlined above. The
resulting expressions are cumbersome and not terribly illuminating.
Instead, one may study the distribution $Q_{\alpha}(Y)$ which obeys
\begin{eqnarray}
\label{QY}
&&Q_\alpha(Y)=q\delta(Y-1)\\
&&+p\int dl\int_0^Y dZ 
{1\over l^{\alpha}}Q_\alpha\left({Z\over l^{\alpha}}\right)
{1\over (1-l)^{\alpha}}Q_\alpha\left({Y-Z\over (1-l)^{\alpha}}\right).\nonumber
\end{eqnarray}
In addition to the recursive nature of the process, we have employed
extensivity, i.e, $\langle Y_\alpha\rangle\propto L^{\alpha}$ in an
interval of length $L$.

Clearly, $Y_0=N$ and $Y_1\equiv 1$, and therefore $Q_1(Y)=\delta(Y-1)$
and $Q_0(N)$ can be determined analytically as well.  Generally,
different behaviors emerge for $\alpha>1$ and $\alpha<1$. We concentrate
on the former case where the support of the distribution $Q_\alpha(Y)$
is the interval [0,1]. The Laplace transform,
$R_\alpha(\lambda)=\int_0^1 dY\,e^{-\lambda Y}Q_\alpha(Y)$, obeys
\begin{equation}
\label{Main}
R_\alpha(\lambda)=q\,e^{-\lambda}+p\int_0^1 dl\,
R_\alpha\left[\lambda l^\alpha\right]   
R_\alpha\left[\lambda (1-l)^\alpha\right].  
\end{equation}
The behavior of $Q_\alpha(Y)$ in the limit $Y\to 0$ is reflected by the
asymptotics of $R_\alpha(\lambda)$ as $\lambda\to \infty$. Substituting
$R_\alpha(\lambda)\sim \exp(-A\lambda^{\beta})$ into both sides of
Eq.~(\ref{Main}), evaluating the integral using steepest decent and
equating the left and right hand sides gives $\beta=1/\alpha$.
Consequently, we find that the distribution has an essential singularity
near the origin
\begin{equation}
\label{Pasymp}
Q_\alpha(Y_\alpha) \sim \exp\left[-BY_\alpha^{-{1\over \alpha-1}}\right], 
\qquad Y_\alpha\to 0.
\end{equation}

Extremal properties can be viewed as an
additional probe of sample to sample fluctuations.  We thus consider
${\cal L}(x)$, the length density of the largest fragment.  For a
self-averaging process with an infinite number of fragments, one expects
${\cal L}(x)\to \delta(x)$ in the thermodynamic limit. To see that
${\cal L}(x)$ is non-trivial for any $p$, let us first determine ${\cal
  L}(x)$ for $x\geq 1/2$.  In this region,
\begin{eqnarray}
\label{Ms}
{\cal L}(x)=q\delta(x-1)+p\int_x^1 {dy\over y}\,
{\cal L}\left({x\over y}\right).
\end{eqnarray}
If the original unit interval has not been fragmented, the largest
fragment is obviously the unit interval.  If the first fragmentation is
performed, only one of the two resulting fragments can be larger than
$x>1/2$.  Therefore, only subsequent breaking of this fragment (of
length $y>x$) can contribute to ${\cal L}(x)$, which explains
Eq.~(\ref{Ms}).

Eq.~(\ref{Ms}) is similar to Eq.~(\ref{pxeq}), and can be solved by the
same technique to give
\begin{eqnarray}
\label{M}
{\cal L}(x)=q\delta(x-1)+pq\,x^{-p}\qquad {\rm for}\quad x\geq 1/2.
\end{eqnarray}
In the complementary case of $x<1/2$, ${\cal L}(x)$ satisfies
\begin{eqnarray}
{\cal L}(x)&=&
p\int\limits_{1-x}^1 {dy\over y}\,{\cal L}\left({x\over y}\right)
+p\int\limits_{1/2}^{1-x} {dy\over y}\,
{\cal L}\left({x\over y}\right){\cal L}_-\left({x\over 1-y}\right)\nonumber\\
&+&p\int\limits_{1/2}^{1-x} {dy\over y}\,
{\cal L}\left({x\over 1-y}\right){\cal L}_-\left({x\over y}\right).
\end{eqnarray}
The first term on the right-hand side of this equation is constructed as
in Eq.~(\ref{Ms}): if we first break the unit interval into two
fragments of lengths $y>1/2$ and $1-y$, then for $1-y<x$ the longest
fragment is produced by breaking the fragment of length $y$.  The next
two terms describe the situation when $1-y>x$, so the longest fragment
can arise out of breaking any of the two fragments.  The factors ${\cal
  L}_-(u) = \int_0^u dv\,{\cal L}(v)$ guarantee that the longest
fragment of length $x$ comes from the fragment of length $v$ in the
first generation. Since ${\cal L}(x)$ obeys different equations in
different regions, it looses analyticity on the boundaries.  Namely,
${\cal L}(x)$ possesses an infinite set of singularities at $x=1/k$
which become weaker as $k$ increases.  Similar singularities underly
extremal properties of a number of random processes including random
walks, spin glasses, random maps, and random trees
\cite{h,djm,der,df,lsp}.

In summary, we have found that recursive fragmentation is scale free,
i.e., the fragment length distribution is purely algebraic.  In higher
dimensions, the volume distribution is a linear combination of $d$
power laws, and consequently, an algebraic divergence characterizes
the small-fragment tail of the distribution. A number of recent impact
fragmentation experiments reported algebraic mass distributions with
the corresponding exponents ranging from $1$ to $2$ \cite{fragm}.  It
will be interesting to further examine whether our simplified model is
appropriate for describing fragmentation of solid objects.

We have also found that the recursive fragmentation process exhibits a
number of features that arise in other complex and disordered systems,
such as non-self-averaging behavior and the existence of an infinite
number of singularities in the distribution of the largest fragment.
These features indicate that even in the thermodynamic limit sample to
sample fluctuations remain, and that knowledge of first order averages
may not be sufficient for characterizing the system. Our 1D model is
equivalent to applying the aforementioned DNA segmentation algorithm to
a random sequence. It will be interesting to study self-averaging and
extremal properties of DNA sequences, which are known to have
commonalities with disordered systems. Indeed, if these subtle features
are found for genetic sequences as well, this would suggest that much
caution should be exercised in statistical analysis of DNA.

\medskip We are thankful to S.~Redner and O.~Weiss for useful
discussions, and to DOE, NSF, ARO, NIH, and DFG for financial support.

\end{multicols}
\end{document}